\documentclass{ws-procs9x6-cpt25}
\begin{document}

\newcommand{\refeq}[1]{(\ref{#1})}
\def\etal {{\it et al.}}
%any other macros go here 
\def\al{\alpha}
\def\be{\beta}
\def\ga{\gamma}
\def\de{\delta}
\def\ep{\epsilon}
\def\ve{\varepsilon}
\def\ze{\zeta}
\def\et{\eta}
\def\th{\theta}
\def\vt{\vartheta}
\def\io{\iota}
\def\vka{\varkappa}
\def\ka{\kappa}
\def\la{\lambda}
\def\vpi{\varpi}
\def\rh{\rho}
\def\vr{\varrho}
\def\si{\sigma}
\def\vs{\varsigma}
\def\ta{\tau}
\def\up{\upsilon}
\def\ph{\phi}
\def\vp{\varphi}
\def\ch{\chi}
\def\ps{\psi}
\def\om{\omega}
\def\Ga{\Gamma}
\def\De{\Delta}
\def\Th{\Theta}
\def\La{\Lambda}
\def\Si{\Sigma}
\def\Up{\Upsilon}
\def\Ph{\Phi}
\def\Ps{\Psi}
\def\Om{\Omega}
\def\cA{{\cal A}}
\def\cB{{\cal B}}
\def\cC{{\cal C}}
\def\cD{{\cal D}}
\def\cE{{\cal E}}
\def\cH{{\cal H}}
\def\cl{{\cal L}}
\def\cL{{\cal L}}
\def\cO{{\cal O}}
\def\cV{{\cal V}}
\def\cP{{\cal P}}
\def\cR{{\cal R}}
\def\cS{{\cal S}}
\def\cT{{\cal T}}

\def\fr#1#2{{{#1}\over{#2}}}
\def\frac#1#2{{\textstyle{{#1}\over{#2}}}}
\def\half{{\textstyle{1\over 2}}}
\def\ol{\overline}
\def\prt{\partial}
\def\pt{\phantom}

\def\Re{\hbox{Re}\,}
\def\Im{\hbox{Im}\,}

\def\lsim{\mathrel{\rlap{\lower4pt\hbox{\hskip1pt$\sim$}}
    \raise1pt\hbox{$<$}}}
\def\gsim{\mathrel{\rlap{\lower4pt\hbox{\hskip1pt$\sim$}}
    \raise1pt\hbox{$>$}}}

\def\etal{{\it et al.}}

\def\vev#1{\langle {#1}\rangle}
\def\expect#1{\langle{#1}\rangle}
\def\bra#1{\langle{#1}|}
\def\ket#1{|{#1}\rangle}

\def\tr{{\rm tr}}

\newcommand{\beq}{\begin{equation}}
\newcommand{\eeq}{\end{equation}}
\newcommand{\bea}{\begin{eqnarray}}
\newcommand{\eea}{\end{eqnarray}}
\newcommand{\rf}[1]{(\ref{#1})}

\def\nn{\nonumber}

\def\psb{\ol\ps{}}

\def\mbf#1{\boldsymbol #1}

\def\Q{\mathcal Q}

\def\pvec{\mbf p}
\def\gavec{\mbf\ga}

\def\Qhat{\widehat\Q}

\def\X{X}
\def\Y{Y}
\def\Z{Z}
\def\Xhat{\widehat\X}
\def\Yhat{\widehat\Y}
\def\Zhat{\widehat\Z}

\def\codt{\cos{\om_\oplus T_\oplus}}
\def\sodt{\sin{\om_\oplus T_\oplus}}
\def\ctodt{\cos{2\om_\oplus T_\oplus}}
\def\stodt{\sin{2\om_\oplus T_\oplus}}

\def\cmtemplate#1#2#3#4{{#1}^{#3}_{#4}}
\def\mfcm#1#2{\cmtemplate{m}{#1}{#2}{5}}
\def\acm#1#2{\cmtemplate{a}{#1}{#2}{}}
\def\bcm#1#2{\cmtemplate{b}{#1}{#2}{}}
\def\ccm#1#2{\cmtemplate{c}{#1}{#2}{}}
\def\dcm#1#2{\cmtemplate{d}{#1}{#2}{}}
\def\ecm#1#2{\cmtemplate{e}{#1}{#2}{}}
\def\fcm#1#2{\cmtemplate{f}{#1}{#2}{}}
\def\gcm#1#2{\cmtemplate{g}{#1}{#2}{}}
\def\Hcm#1#2{\cmtemplate{H}{#1}{#2}{}}

\def\ctemplate#1#2#3#4{{#1}^{(#2)#3}_{#4}}
\def\mc#1#2{\ctemplate{m}{#1}{#2}{}}
\def\mfc#1#2{\ctemplate{m}{#1}{#2}{5}}
\def\ac#1#2{\ctemplate{a}{#1}{#2}{}}
\def\bc#1#2{\ctemplate{b}{#1}{#2}{}}
\def\cc#1#2{\ctemplate{c}{#1}{#2}{}}
\def\dc#1#2{\ctemplate{d}{#1}{#2}{}}
\def\ec#1#2{\ctemplate{e}{#1}{#2}{}}
\def\fc#1#2{\ctemplate{f}{#1}{#2}{}}
\def\gc#1#2{\ctemplate{g}{#1}{#2}{}}
\def\Hc#1#2{\ctemplate{H}{#1}{#2}{}}

\def\mcf#1#2{\ctemplate{m}{#1}{#2}{F}}
\def\mfcf#1#2{\ctemplate{m}{#1}{#2}{5F}}
\def\acf#1#2{\ctemplate{a}{#1}{#2}{F}}
\def\bcf#1#2{\ctemplate{b}{#1}{#2}{F}}
\def\ccf#1#2{\ctemplate{c}{#1}{#2}{F}}
\def\dcf#1#2{\ctemplate{d}{#1}{#2}{F}}
\def\ecf#1#2{\ctemplate{e}{#1}{#2}{F}}
\def\fcf#1#2{\ctemplate{f}{#1}{#2}{F}}
\def\gcf#1#2{\ctemplate{g}{#1}{#2}{F}}
\def\Hcf#1#2{\ctemplate{H}{#1}{#2}{F}}

\def\mcpf#1#2{\ctemplate{m}{#1}{#2}{\prt F}}
\def\mfcpf#1#2{\ctemplate{m}{#1}{#2}{5\prt F}}
\def\acpf#1#2{\ctemplate{a}{#1}{#2}{\prt F}}
\def\bcpf#1#2{\ctemplate{b}{#1}{#2}{\prt F}}
\def\ccpf#1#2{\ctemplate{c}{#1}{#2}{\prt F}}
\def\dcpf#1#2{\ctemplate{d}{#1}{#2}{\prt F}}
\def\ecpf#1#2{\ctemplate{e}{#1}{#2}{\prt F}}
\def\fcpf#1#2{\ctemplate{f}{#1}{#2}{\prt F}}
\def\gcpf#1#2{\ctemplate{g}{#1}{#2}{\prt F}}
\def\Hcpf#1#2{\ctemplate{H}{#1}{#2}{\prt F}}

\def\cmtemplate#1#2#3#4{{#1}^{#3}_{#4}}
\def\mfcmw#1#2#3{\cmtemplate{m}{#1}{#2}{5{,#3}}}
\def\acmw#1#2#3{\cmtemplate{a}{#1}{#2}{{#3}}}
\def\bcmw#1#2#3{\cmtemplate{b}{#1}{#2}{{#3}}}
\def\ccmw#1#2#3{\cmtemplate{c}{#1}{#2}{{#3}}}
\def\dcmw#1#2#3{\cmtemplate{d}{#1}{#2}{{#3}}}
\def\ecmw#1#2#3{\cmtemplate{e}{#1}{#2}{{#3}}}
\def\fcmw#1#2#3{\cmtemplate{f}{#1}{#2}{{#3}}}
\def\gcmw#1#2#3{\cmtemplate{g}{#1}{#2}{{#3}}}
\def\Hcmw#1#2#3{\cmtemplate{H}{#1}{#2}{{#3}}}

\def\ctemplate#1#2#3#4{{#1}^{(#2)#3}_{#4}}
\def\mcw#1#2#3{\ctemplate{m}{#1}{#2}{{#3}}}
\def\mfcw#1#2#3{\ctemplate{m}{#1}{#2}{5{,#3}}}
\def\acw#1#2#3{\ctemplate{a}{#1}{#2}{{#3}}}
\def\bcw#1#2#3{\ctemplate{b}{#1}{#2}{{#3}}}
\def\ccw#1#2#3{\ctemplate{c}{#1}{#2}{{#3}}}
\def\dcw#1#2#3{\ctemplate{d}{#1}{#2}{{#3}}}
\def\ecw#1#2#3{\ctemplate{e}{#1}{#2}{{#3}}}
\def\fcw#1#2#3{\ctemplate{f}{#1}{#2}{{#3}}}
\def\gcw#1#2#3{\ctemplate{g}{#1}{#2}{{#3}}}
\def\Hcw#1#2#3{\ctemplate{H}{#1}{#2}{{#3}}}

\def\mcfw#1#2#3{\ctemplate{m}{#1}{#2}{F{,#3}}}
\def\mfcfw#1#2#3{\ctemplate{m}{#1}{#2}{5F{,#3}}}
\def\acfw#1#2#3{\ctemplate{a}{#1}{#2}{F{,#3}}}
\def\bcfw#1#2#3{\ctemplate{b}{#1}{#2}{F{,#3}}}
\def\ccfw#1#2#3{\ctemplate{c}{#1}{#2}{F{,#3}}}
\def\dcfw#1#2#3{\ctemplate{d}{#1}{#2}{F{,#3}}}
\def\ecfw#1#2#3{\ctemplate{e}{#1}{#2}{F{,#3}}}
\def\fcfw#1#2#3{\ctemplate{f}{#1}{#2}{F{,#3}}}
\def\gcfw#1#2#3{\ctemplate{g}{#1}{#2}{F{,#3}}}
\def\Hcfw#1#2#3{\ctemplate{H}{#1}{#2}{F{,#3}}}

%new definitions
\def\bpftw#1#2{{\widetilde b}_{\prt F,#1}^{#2}}
\def\bcpfw#1#2#3{\ctemplate{b}{#1}{#2}{\prt F{,#3}}}
\def\Hcpfw#1#2#3{\ctemplate{H}{#1}{#2}{\prt F{,#3}}}

\def\mn{{\mu\nu}}
\def\ma{{\mu\al}}
\def\mna{{\mu\nu\al}}
\def\ab{{\al\be}}
\def\bec{{\be\ga}}
\def\mab{{\mu\al\be}}
\def\mnab{{\mu\nu\al\be}}
\def\abc{{\al\be\ga}}
\def\bca{{\be\ga\al}}
\def\cab{{\ga\al\be}}
\def\mabc{{\mu\al\be\ga}}
\def\mnabc{{\mu\nu\al\be\ga}}
\def\abcd{{\al\be\ga\de}}
\def\va{{\vs\al}}
\def\vab{{\vs\al\be}}
\def\vabc{{\vs\al\be\ga}}
\def\vabcd{{\vs\al\be\ga\de}}

\def\m{m_\ps}
\def\mw{m_w}
\def\Z{\si}

\def\quar{\frac 1 4}

\def\Epn{\cE_{n,\pm1}^{e^-}}
\def\En{E_{n,\pm1}^{e^-}}
\def\app{\approx}
\def\cthodt{\cos{3\om_\oplus T_\oplus}}
\def\sthodt{\sin{3\om_\oplus T_\oplus}}
\def\note#1{{\it note \cite{#1}}}
\def\ens{E_{n,s}}
\def\enms{E_{n,-s}}
\def\at{\widetilde a}
\def\bt{\widetilde b}
\def\bft{{\widetilde b}_F}
\def\mft{{\widetilde m}_F}
\def\atw#1#2{{\widetilde a}_{#1}^{#2}}
\def\btw#1#2{{\widetilde b}_{#1}^{#2}}
\def\bftw#1#2{{\widetilde b}_{F,#1}^{#2}}
\def\mftw#1#2{{\widetilde m}_{F,#1}^{#2}}
\def\atws#1#2{{\widetilde a}_{#1}^{*#2}}
\def\btws#1#2{{\widetilde b}_{#1}^{*#2}}
\def\bftws#1#2{{\widetilde b}_{F,#1}^{*#2}}
\def\mftws#1#2{{\widetilde m}_{F,#1}^{*#2}}
%new definitions
\def\widecheck#1{\hskip#1pt\huge$\check{}$}
\def\bighacek#1#2{\vbox{\ialign{##\crcr\widecheck#2\crcr
			\noalign{\kern-10pt\nointerlineskip}
			$\hfil\displaystyle{#1}\hfil$\crcr}}}
\def\hb{\bighacek{b}{3}}
\def\bfhb#1#2{\hb{}_{F,#1}^{#2}}
\def\bfhbs#1#2{\hb{}_{F,#1}^{*#2}}
\def\bpfhb#1#2{\hb{}_{\prt F,#1}^{#2}}
\def\bpfhbs#1#2{\hb{}_{\prt F,#1}^{*#2}}
\def\hbb{\bighacek{\mbf b}{3}}
\def\bbfhb#1#2{\hbb{}_{F,#1}^{#2}}
\def\bbfhbs#1#2{\hbb{}_{F,#1}^{*#2}}

\title{Lorentz and CPT Tests
in Neutron and Storage-Ring EDM Experiments}

\author{Yunhua Ding}

\address{Department of Physics and Astronomy, 
Ohio Wesleyan University, \\
Delaware, OH 43015, USA}

\begin{abstract}
We investigate Lorentz- and CPT-violating effects in neutron and storage-ring electric dipole moment (EDM) experiments within the framework of the Standard-Model Extension (SME).
For neutron EDM experiments, perturbation theory is applied to derive leading-order contributions to the spin precession frequency arising from Lorentz and CPT violation. 
For storage-ring experiments, a generalized Bargmann-Michel-Telegdi equation is used to determine the corresponding spin-precession modifications. 
The analysis establishes explicit correspondences between measured EDMs and specific SME coefficients, 
providing a basis for setting the first limits on several previously unconstrained coefficients for Lorentz violation in future studies.

\end{abstract}

\section{Introduction}
\label{introduction}

Electric dipole moment experiments provide one of the most sensitive probes of possible new sources of $P$ and $CP$ violation in nature.
For example,
the neutron EDM (nEDM) has been measured with a precision of
$|d_n|<1.8 \times 10^{-26}~e$~cm (90\% C.L.),
\cite{20abel}
while measurements of the muon EDM ($\mu$EDM)
yield $|d_{\mu^+}|<2.1\times 10^{-19}~e$~cm for antimuons, and $|d_{\mu^-}|<1.5\times 10^{-19}~e$~cm for muons (95\% C.L.).
\cite{09ben}
The impressive sensitivities achieved in these measurements also make EDM experiments powerful tests of other fundamental symmetries, including Lorentz and CPT symmetry.
It has been shown that tiny violations of Lorentz and CPT symmetry can arise in certain candidate theories unifying gravity with quantum physics, such as string theory.
\cite{string}
A comprehensive framework to describe all possible Lorentz-violating effects within effective field theory is provided by the Standard-Model Extension,
constructed by adding all possible Lorentz-violating terms into the action of General Relativity and the Standard Model.
\cite{SME}
Each Lorentz-violating term involves a coordinate-independent contraction of a Lorentz-violating operator with a corresponding controlling coefficient.
In effective field theory, any CPT violation is necessarily accompanied by Lorentz violation,
so the SME also encompasses all possible CPT-violating effects.
\cite{owg}
Extensive high-precision tests across a wide range of subfields have placed stringent limits on many SME coefficients.
\cite{datatables}

In this work, we analyze Lorentz- and CPT-violating contributions to spin precession in both neutron and storage-ring EDM experiments
within the SME framework.
For confined neutrons, perturbation theory is used to determine the leading-order effects.
For charged particles in storage rings, we apply the generalized Bargmann-Michel-Telegdi (BMT) equation
to extract the corresponding spin-precession modifications.
\cite{25dkv}
We establish explicit relations between measured EDMs and specific SME coefficients.
This enables future studies to place first constraints on various SME coefficients using current EDM limits reported by neutron and storage-ring experiments.   
\cite{20abel,09ben}

\section{Neutron EDM}
\label{neutron EDM}

In conventional Lorentz-invariant analyses of EDM effects involving confined neutrons,
the primary observable is the shift in the neutron spin-precession frequency,
corresponding to the energy difference between spin-up and spin-down states.
To isolate the EDM contribution, experiments measure the spin-precession frequency with parallel and antiparallel electric and magnetic fields.
The resulting difference is proportional to the nEDM according to
\bea
\label{Deomedm}
\De \om_{n, \rm EDM} = 4 d_n E,
\eea
where $d_n$ is the nEDM and $E$ is the applied electric-field strength.

Lorentz violation can induce frequency shifts even when the nEDM vanishes.
The dominant energy shifts for neutrons in this case can be obtained via perturbation theory as $\de E_s = \vev{\ch_s | \de \cH | \ch_s}$,
where $\ch_s$ are unperturbed stationary neutron eigenstates with $s=\pm 1$ denoting spin orientation relative to the magnetic field,
and $\de \cH$ is the perturbative Hamiltonian for Lorentz violation.
\cite{16dk}
For convenience,
we adopt an apparatus frame with magnetic field
$\mbf B = B \hat x_3$ and electric field
$\mbf E = E \hat x_3$.
In this frame, reversing the electric-field direction while holding the magnetic field fixed yields a Lorentz-violating frequency difference
\bea
\label{Deomlv}
\De \om_{n, \rm LV} = 4 \bftw n {303} E,
\eea
where the tilde coefficient $\bftw n {303}$ is
\bea
\bftw n {303} = \bcfw 5 {303} n + \Hcfw 5 {1203} n - m_n \dcfw 6 {3003} n - m_n \gcfw 6 {12003} n.
\eea
Comparison of Eqs.~\rf{Deomedm} and \rf{Deomlv} reveals the correspondence
\bea
\label{nedmcorres}
|d_n| \longleftrightarrow |\bftw n {303}|,
\eea
relating the measured nEDM to the relevant coefficients for Lorentz violation.

\section{Storage-Ring EDM}
\label{storage-ring EDM}

Precision EDM measurements of charged particles,
such as protons, muons, and their antiparticles,
are performed in storage-ring experiments,
where particles are confined by electric and/or magnetic fields,
and the EDM-induced spin precession is measured
via polarimetry from spin-dependent elastic scattering along the ring.
The spin-precession frequency in the presence of an EDM is described by the Bargmann-Michel-Telegdi (BMT) equation.
\cite{bmt}
For a radially outward electric field $\vec{E}$
and a uniform upward magnetic field $\vec{B}$,
the lab-frame EDM contribution to the spin-precession frequency for a particle of species $w$, EDM $d_w$, and velocity $\vec{\be}$ is
\bea
\label{omedm}
\de \vec{\om}_{s, \rm EDM} = -2 d_w (\vec{E} + \vec{\be} \times \vec{B}).
\eea
Recently,
the BMT equation was generalized to incorporate Lorentz and CPT violation.
\cite{25dkv}
From the generalized BMT equation,
the modification to the radial component of the spin-precession frequency is
\bea
\label{omhacheck}
\de \om^{\rm rad}_{s, \rm {LV}}
&=&
\bighacek{H}{3}_{w}^{03}
+
E (\bighacek{b}{3}_{F,w}^{11}
+ \bighacek{b}{3}_{F,w}^{22}
+\bighacek{g}{3}_{F,w}^{33})
+
B \bighacek{H}{3}_{F, w}^{33},
\eea
where the hacheck coefficients are defined as
\bea
\bighacek{H}{3}_{w}^{03} 
&=&
-\beta \Big( 2 H^{0 3} - m_w (d^{1 2} - d^{2 1}) 
- \gamma m_w (2 g^{0 3 0} + g^{1 3 1} + g^{2 3 2}) \Big),
\nn\\
\bighacek{b}{3}_{F,w}^{11} 
&=&
\frac{1}{\gamma} b_F^{1 0 1} + H_F^{2 3 0 1} - m_w d_F^{1 0 0 1} - \gamma m_w g_F^{2 3 0 0 1},
\nn\\
\bighacek{b}{3}_{F,w}^{22} 
&=&
\frac{1}{\gamma} b_F^{2 0 2} - H_F^{1 3 0 2} - m_w d_F^{2 0 0 2} + \gamma m_w g_F^{1 3 0 0 2},
\nn\\
\bighacek{g}{3}_{F,w}^{33} 
&=&
\gamma \beta^2 m_w ( g_F^{0 3 1 0 2} - g_F^{0 3 2 0 1} ),
\nn\\
\bighacek{H}{3}_{F, w}^{33} 
&=&
2 \beta H_F^{0 3 1 2} - \beta m_w ( d_F^{1 2 1 2} - d_F^{2 1 1 2} )
- \gamma \beta m_w ( 2 g_F^{0 3 0 1 2} + g_F^{1 3 1 1 2} + g_F^{2 3 2 1 2} )
\nn\\
\eea
with $m_w$ and $\ga=1/\sqrt{1-\be^2}$ indicating the particle's mass and Lorentz factor, respectively.
Comparison of Eqs.~\rf{omedm} and \rf{omhacheck} yields the correspondence between a measured EDM and the relevant SME coefficients,
\bea
\label{sredmcorres}
|d_{w}| \longleftrightarrow \fr{|\de \om^{\rm rad}_{s, \rm {LV}}|}{2|\vec{E} + \vec{\be} \times \vec{B}|}.
\eea

\section{Summary} 
\label{summary}

The relations in Eqs.~\rf{nedmcorres} and \rf{sredmcorres} establish direct correspondences between EDM observables and the relevant SME coefficients for Lorentz violation in both neutron and storage-ring experiments.
Placing bounds on the relevant SME coefficients in the Sun-centered frame requires transformations from the 
apparatus frame to the Sun-centered frame,
as discussed in detail elsewhere.
\cite{16dk}
Existing experimental limits on the neutron EDM and on the EDMs of muons and antimuons,
\cite{20abel, 09ben} 
can therefore be translated into new constraints on a variety of previously unbounded SME coefficients,
an analysis that can be performed in future work.

\section{Acknowledgments}
This work is supported in part by the Teaching, Learning, and Innovation
grant from Ohio Wesleyan University.

\end{document}